\documentclass[%
 reprint,
nofootinbib,
 amsmath,amssymb,
 aps,
]{revtex4-2}

\usepackage{graphicx}% Include figure files
\usepackage{dcolumn}% Align table columns on decimal point
\usepackage{bm}% bold math
\usepackage{hyperref}% add hypertext capabilities
\usepackage{orcidlink}

\begin{document}

\preprint{APS/123-QED}

\title{Approximating the statistics of a gravitational wave background}% Force line breaks with \\

\author{Mikel Falxa~\orcidlink{0000-0002-0815-1781}}
\thanks{mikel.falxa@dipc.org}
\affiliation{Donostia International Physics Center (DIPC),\\Paseo Manuel de Lardizabal 4, 20018 Donostia-San Sebastian, Spain}%Lines break automatically or can be forced with \\

% \date{\today}% It is always \today, today,
             %  but any date may be explicitly specified

\begin{abstract}
The astrophysical origin of the gravitational wave background (GWB) reported by pulsar timing array (PTA) collaborations has yet to be confirmed. A GWB signal made of the sum of individual gravitational wave (GW) from the population of supermassive black hole binaries (SMBHB) would show the imprint of a discrete Poissonian statistics in its spectral properties. In this work, we propose a tool based on the saddlepoint approximation method to estimate the distribution of characteristic strain for any given population model. This tool can be used for Bayesian inference or for a quick visualization of the statistics of the GWB from the output of more realistic semi-analytical models. We introduce the general family of variance mixture Gaussian distributions that models heavy-tailed behavior distributions that is expected for a non-Gaussian GWB signal. We show that setting the correct hierarchical priors to a Gaussian free spectrum PTA likelihood is effectively equivalent to a non-Gaussian likelihood. Using ideal simulations, we compare the performance of the saddlepoint approximation with a log-Normal distribution to infer the parameters of the astrophysical model from the statistics of the GWB and show that correctly modeling higher order moments is essential. Future PTA analyses should include the statistics of the GWB in their pipelines.
\end{abstract}

%\keywords{Suggested keywords}%Use showkeys class option if keyword
                              %display desired
\maketitle

%\tableofcontents

\section{Introduction}

A gravitational wave background (GWB) is a stochastic signal that emerges from the superposition of many individual gravitational wave (GW) signals \cite{phinney2001}. The sources of these GWs are often compact object binary systems that produce the loudest GWs in the Universe. These sources are usually part of a population that follow statistical properties which carry interesting astrophysical information about their formation and distribution. Logically, the population dictates the amplitude and spectral properties of the stochastic GWB \cite{holodeck}. Thus, it should be possible to infer some of the astrophysical population parameters through the measurement of a GWB signal.

In Pulsar timing arrays (PTA), the GWB is a nanohertz red noise showing up in the precision timing of pulsars \cite{Sazhin, Detweiler}, produced by the population of supermassive black holes binaries (SMBHBs) in the Universe that are formed through galaxy merger \cite{sesana2009, Sesana_2008}. Evidence for such a signal in PTA data was reported some years ago \cite{epta_wm3, ng15yr, cpta, ppta_dr3, mpta}. The distribution of SMBHBs in mass, redshift and frequency is governed by galaxy dynamics and supermassive black hole (SMBH) growth history \cite{bonoli2025}. It is then natural to consider a hierarchical framework where the properties of the measured GWB are modeled by more fundamental astrophysical parameters. This approach was already adopted in previous studies \cite{goncharov2026, satopolito_2024, xue2025} where the authors evaluate the expected statistics of a GWB signal to connect it to the PTA statistical framework. This statistics depends heavily on the number and properties of the GW sources making up the GWB. Specifically, the discreteness of the population induces non-Gaussian behaviors that contain some information about the distribution of sources. When a few loud sources dominate the signal, the central limit theorem does not apply and the GWB shows strong signs of non-Gaussianity. This is what is expected for a realistic population of SMBHBs. The non-Gaussianity of the signal has been extensively discussed and investigated in recent works. It was already discussed in \cite{xue2025} where the authors developped a framework based on compound poisson processes to model the GWB signal. In \cite{satopolito_2025}, a similar approach is used to estimate and infer the spectral distribution the GWB to infer parameters. In \cite{yacine} an analytical expression for the statistics of the GWB is proposed. More recently, \cite{raidal2026} investigated the heavy tail behavior of the GWB statistics, and \cite{goncharov2026} proposes a way to split the GWB and single loudest source statistics in a joint hierarchical framework. Additionally, the non-Gaussian statistics of the GWB might be hard to detect in a purely agnostic way, as explained in \cite{cecchini2026}, due to the very broad response of PTAs to GWs. Other studies propose to detect them with higher order statistics using the four point correlator \cite{kuntz2026}, or setting the correct non-Gaussian priors on the Fourier decomposition of the signal in Bayesian analysis \cite{falxa_ng, gundersen2026}.

In this work, we propose a method based on the saddlepoint approximation to estimate the statistics of the GWB spectrum. This method is already discussed in the appendix of \cite{goncharov2026}. Here, we propose a code available at \url{https://github.com/mfalxa/zamari} that has the purpose of being flexible in terms of astrophysical modeling. It is designed to accept any distribution of sources in chirp mass and redshift as well as binary frequency evolution. It could be paired with outputs of semi-analytical models to estimate the expected statistics of the GWB. The saddlepoint method offers a reasonable approximation to the distribution of characteristic strain without requiring heavy training on numerous simulations. Still, machine learning methods as developped in \cite{laal2024, Bonetti_2024} can present advantages, specifically to account for the eccentricity of the binaries, that is discarded in this work and must be accounted for in future works.

This article is organized as follows. In the first section, we review the approximations that lead to the common expression of the GWB statistics in the Gaussian ensemble approximation. In the next section, we present the family of heavy tailed scale mixture distributions that can be constructed from Gaussian distributions. Then, we present the saddlepoint approximation method and how we can use it for the GWB, after what we introduce the phenomenological model and the Bayesian framework used in this work. Finally, we share the results that are obtained for high mass and low mass binary populations, comparing the performance of the saddlepoint approximation with a log-normal distribution.

\section{Gravitational wave background}

The timing residuals $\delta t_a$ measured in pulsar $a$ for a GWB signal consist of the sum of $N_s$ individual GW signals with amplitude $A_j$, frequency $f_j$, and phase $\Phi_j$ as \cite{raidal2026, Lamb_2026}

\begin{equation}
    \delta t_a = \sum_{j=1}^{N_S} \frac{A_j}{4\pi if_j} R_{j,a} e^{i2\pi f_j t + \phi_j} + c.c. ,
\end{equation}
where $R_{a,j}$ is the response of pulsar $a$ to the GW signal emitted by source $j$, given by

\begin{equation}
\begin{aligned}
    R_{j,a} & = \left [1 - e^{-2i\pi f_j L_a (1 + \hat \Omega_j \cdot \hat p_a )}\right]\\
    & \times \left[ \frac{1 + \cos ^2 \iota_j}{2} F_a^+ - i \cos \iota_j F_a ^\times \right],
\end{aligned}
\end{equation}
with $L_a$ the distance Earth-pulsar $a$, $\iota_j$ the inclination of source $j$, $\hat \Omega_j$ the unit vector pointing to the sky location of source $j$ and $F^+_a, F^\times_a$ the antenna pattern response for pulsar $a$ corresponding to polarization modes $+$ and $\times$ of the GW, defined as

\begin{equation}
    F^A_a (\hat \Omega_j, \psi_j) = \frac{1}{2} \frac{\hat p_a^\alpha \hat p_a^\beta}{1 + \hat \Omega_j \cdot \hat p_a} \epsilon^A_{\alpha \beta} (\hat \Omega_j, \psi_j),
\end{equation}
where $\epsilon^A_{\alpha \beta}$ is the polarization tensor that can be rotated along the polarization basis by an angle $2\psi$ as

\begin{equation}
\begin{bmatrix}
\epsilon^+_{\alpha\beta} \\
\epsilon^\times_{\alpha\beta}
\end{bmatrix}
\rightarrow
\begin{bmatrix}
\cos(2\psi) & -\sin(2\psi) \\
\sin(2\psi) & \cos(2\psi)
\end{bmatrix} 
\begin{bmatrix}
\epsilon^+_{\alpha\beta} \\
\epsilon^\times_{\alpha\beta}
\end{bmatrix}.
\end{equation}

The timing residuals expressed in Fourier domain are given by

\begin{equation}
    \delta \tilde t_a (f) = \sum_{j=1}^{N_S} \frac{A_j}{4\pi if_j} \left [ R_{j,a} e^{i\phi_j} w_{j}^{+} (f) - R_{j,a}^*e^{-i\phi_j} w_{j}^{-} (f) \right].
\end{equation}

The finite total time of observation $T$ produces spectral leakage modelled by the functions $w_{j}^{\pm} (f, f_s)$ \cite{crisostomi2025, raidal2026, Lamb_2026}. In all rigor, because of the unevenly sampled nature of PTA data, pulsars can have different total time $T_a$ and effective cadence of observation, so these functions are not exactly the same and should wear an index $a$. But in PTA data analysis, this effect is usually neglected and we use the approximation $w_{j}^{\pm} (f, f_s) \approx \delta (f \pm f_s)$. It is important to note that the latter can have consequences on the inference \cite{crisostomi2025, Quelquejay_Leclere_2026}.

We can replace the discrete sum by an integral weighted by the amplitude map $A(\hat \Omega, f_s, \psi, \iota)$ as

\begin{equation}
\begin{aligned}
    \delta \tilde t_a (f) & \equiv \int df_s\int \frac{d \hat \Omega}{4\pi} \int \frac{d\psi}{\pi} \int \frac{d\cos \iota}{2}\frac{A(\hat \Omega, f_s, \psi, \iota)}{4\pi i f_s}\\
    & \times \left [R_a (\hat \Omega, f_s, \psi, \iota) e^{i\Phi(\hat \Omega, f_s)} w^{+} (f, f_s) \right.\\
    & \left. - R^*_a (\hat \Omega, f_s, \psi, \iota) e^{-i\Phi(\hat \Omega, f_s)} w^{-} (f, f_s) \right].
\end{aligned}
\end{equation}

Since individual binaries are independent and their phase uniformly distributed, we have the expected value $\langle e^{i(\Phi(\hat \Omega, f_s) - \Phi(\hat \Omega, f_s))}\rangle = \delta (\hat \Omega - \hat \Omega') \delta(f_s - f'_s)$ and $\langle e^{i\Phi(\hat \Omega, f_s)}\rangle = 0$ where $\langle . \rangle$ denotes the ensemble average. This gives the cross correlated timing residuals

\begin{equation}
\begin{aligned}
    \langle \delta \tilde t_a ^* (f) \delta \tilde t_b (f) \rangle & = \int df_s\int \frac{d\hat\Omega}{4\pi} \frac{\langle A^2(\hat \Omega, f_s)\rangle}{16\pi^2 f_s^2} \langle R^*_a(\hat \Omega) R_b(\hat \Omega) \rangle_{\psi, \iota}\\
    & \times \left [ (w^+)^2(f, f_s) + (w^-)^2(f, f_s)\right].
\end{aligned}
\end{equation}

In this last step, we assumed that polarization and inclination are uniformly distributed for all sources by setting $A(\hat \Omega, f_s, \psi, \iota) \equiv A(\hat \Omega, f_s)$, yielding the polarization and inclination averaged cross response $\langle R^*_a(\hat \Omega) R_b(\hat \Omega) \rangle_{\psi, \iota}$. Note that we also dropped the frequency dependence of the response function, which is a reasonable approximation in the long detector arm limit (see Appendix~\ref{app:orf}).

The squared amplitude map $A^2(\hat \Omega, f_s)$ can be decomposed into spherical harmonics as $A^2(\hat \Omega, f_s) = \sum_{lm} c_{lm}(f_s) Y_{lm}(\hat \Omega)$ to model the anisotropic power distribution of the GWB \cite{Taylor_2020}. In the large number of source limit, the main contributor is the monopole $c_{00}$ and the background is considered isotropic, meaning that $A^2(\hat \Omega, f_s) \approx A^2(f_s) = c_{00} (f_s)$\footnote{There is a nuance here. Technically, if one considers statistical isotropy, then the average Universe across many realizations is isotropic, and $c_{00}$ is the only remaining term. However, for higher order moments, the $c_{lm}$ with $l>0$ do not vanish, because fluctuations at smaller angular scale are present due to the discreteness of the population. This is why anisotropies and non-Gaussian statistics are closely related.}. This approximation is no longer valid when single binaries dominate the signal, as it is expected when measuring one realization of the Universe. A single binary dominated signal would also break the assumption of unpolarized GWB.

Defining
\\
\begin{equation}
\begin{aligned}
\Gamma_{00,ab} & =\int \frac{d\hat\Omega}{4\pi} \langle R^*_a(\hat \Omega) R_b(\hat \Omega)\rangle_{\psi, \iota} = \frac{2}{3} \textrm{HD} (\zeta_{ab}),\\
\frac{c_{00}^W (f)}{8\pi^2 f^2} & = \int df_s \frac{c_{00}(f_s)}{16\pi^2 f_s^2} \left [ (w^+)^2(f, f_s) + (w^-)^2(f, f_s)\right],
\end{aligned}
\end{equation}
with $\textrm{HD} (\zeta_{ab})$ the Hellings-Downs correlations \cite{HD} between pulsars $a$ and $b$ (see Appendix~\ref{app:orf}), and $c_{00}^W$ the convoluted amplitude with $w_{j}^{\pm} (f, f_s)$, proportional to the one-sided power spectral density (PSD). Finally, we get a simple expression for the cross correlated timing residuals between pulsars $a$ and $b$

\begin{equation}
\begin{aligned}
\langle \delta \tilde t_a ^* (f) \delta \tilde t_b (f) \rangle = \frac{\langle c_{00}^W (f) \rangle}{12\pi^2 f^2} \textrm{HD} (\zeta_{ab}).
\end{aligned}
\label{eq:cross_correlated_residuals}
\end{equation}

In the Gaussian ensemble approximation, $\delta \vec t \sim \mathcal{N}(0, C_{ab})$ with $C_{ab}$ that is defined by the pair correlations $\langle \delta \tilde t_a ^* (f) \delta \tilde t_b (f) \rangle$. This approximation appears to have limited consequences on the inference of astrophysical parameters, as shown in \cite{xue2025, raidal2026}, when accounting for the fluctuation of the number of sources. We dedicate a more in-depth discussion related to this approximation in Appendix ~\ref{app:gaussian_approx}.

Then, defining $X, Y \sim \mathcal{N}(0,\textrm{HD} (\zeta_{ab}))$, we have \footnote{Using $\langle X^2 + Y ^2 \rangle = 2\textrm{HD} (\zeta_{ab})$ and $\langle X^4 + Y^4 + 2 X^2 Y^2 \rangle = 8\textrm{HD} (\zeta_{ab})^2$}

\begin{equation}
   \delta \tilde t_a (f) \sim \sqrt{\frac{c^W_{00} (f)}{12\pi^2 f^2}}  \times \frac{X + iY}{\sqrt{2}},
\end{equation}

\begin{equation}
\begin{aligned}
   \langle \delta \tilde t^*_a (f) \delta \tilde t_b (f) \rangle = \frac{ \langle c^W_{00} (f) \rangle}{12 \pi^2 f^2} \textrm{HD} (\zeta_{ab}),
\end{aligned}
\end{equation}

\begin{equation}
\begin{aligned}
   \langle (\delta \tilde t^*_a (f) \delta \tilde t_b (f))^2 \rangle = \frac{\langle (c^W_{00} (f))^2 \rangle}{72\pi^4 f^4}  \textrm{HD}^2 (\zeta_{ab}),
\end{aligned}
\end{equation}
so higher order moments depend on the statistics of $c_{00}^W$.

When considering one realization of the Universe, $c_{00}$ is fixed. However, accounting for the source population statistics (i.e. the cosmic variance) requires that $c_{00}$ follows a certain probability distribution given by the astrophysical population parameters. Therefore, even in the assumption of Gaussian distributed timing residuals, the full distribution is in fact non-Gaussian, of the form $\delta t \sim \sqrt{V} \times \mathcal{N}(0, 1)$ where $V \sim g_{v}$. This type of distribution belongs to the family of heavy tailed distributions, the Normal variance-mean mixture distributions.

\section{Normal variance-mean mixture distributions}
\label{sec:NVM}

In its general form, a normal variance-mean mixture (NVM) random variable is defined as \cite{nvm, yu_nvm}

\begin{equation}
    X = \alpha + \beta V + \sqrt{V} Z,
\end{equation}
where $V \sim g(v)$ is a probability distribution function of parameters $\lambda$ and $Z\sim \mathcal{N}(0, 1)$ with a distribution defined as

\begin{equation}
    f(x) = \int_0 ^\infty dv \frac{1}{\sqrt{2\pi v}} \exp \left\{-\frac{(x -\alpha - \beta v)^2}{2v} \right\} g(v).
\end{equation}

In the symmetric and centered case, $\alpha, \beta = 0$.

\begin{equation}
\begin{aligned}
    f(x) & = \int_0 ^\infty dv \frac{1}{\sqrt{2\pi v}} \exp \left\{-\frac{x^2}{2v} \right\} g(v)\\
    & \equiv  \int_0 ^\infty dv \phi(x|v) g(v),
\end{aligned}
\label{eq:nongaussian_prior}
\end{equation}
which can be interpreted as the continuous equivalent of a Gaussian mixture model \cite{falxa_ng} (the integral replaces the sum) or, in Bayesian terms, the marginal distribution of $\phi(x|v) g(v)$ where $g(v)$ is the prior on the variance of the Gaussian distribution.

For the symmetric case $X = \sqrt{V} Z$, the mean and skewness are zero while the variance $\sigma_X^2$ and excess kurtosis $\kappa_X$ are given by

\begin{equation}
    \sigma_X^2 = \langle V \rangle \langle Z^2 \rangle = \mu_V,
\end{equation}

\begin{equation}
    \kappa_X = \left( \langle V^2 \rangle - \langle V \rangle^2 \right ) \langle Z^4 \rangle = 3 \sigma^2_V.
\label{eq:match_kurtosis}
\end{equation}

This result can be easily generalized to multivariate distributions considering that $Z$ is a multivariate normal and using Isserlis's theorem to find higher order moments, as we show in \autoref{sec:param_NG}. Thus, we can construct families of non-Gaussian distributions for different $g(v)$ and control the first two non zero cumulants $\sigma_X^2$ and $\kappa_X$ by matching the mean $\mu_V$ and variance $\sigma^2_V$ of $V$ to the parameters of the distribution. Here we list some examples for which the matching is possible analytically.

\begin{itemize}
    \item Log-normal distribution,
    \begin{equation}
        g(v|\mu_V, \sigma_V) = \frac{1}{v\sigma\sqrt{2\pi}} \exp \left \{ - \frac{(\ln v - \mu)^2}{2 \sigma^2} \right \},
    \end{equation}
    with $\mu = \ln \left( \frac{\mu_V^2}{\sqrt{\mu_V^2 + \sigma_V^2}} \right)$ and $\sigma = \sqrt{ \ln \left ( 1 + \frac{\sigma^2_V}{\mu^2_V}\right)}$, which we use in this work.
    
    \item Inverse Gaussian distribution,
        \begin{equation}
        g(v|\mu_V, \sigma_V) = \sqrt{\frac{\lambda}{2\pi v^3}} \exp \left \{ - \frac{\lambda (v - \mu)}{2 \mu^2 v} \right \},
    \label{eq:ig}
    \end{equation}
    with $\mu = \mu_V$ and $\lambda = \mu_V^3 / \sigma^2_V$. The NVM obtained from the generalized inverse Gaussian distribution produces generalized hyperbolic distributions as used in \cite{karnesis_2025}.
    
    \item Gamma distribution,
    \begin{equation}
        g(v|\mu_V, \sigma_V) = \frac{1}{\Gamma (\alpha) \theta^\alpha} v^{\alpha - 1} \exp \left \{ - \frac{v}{\theta} \right \},
    \end{equation}
    with $\alpha = \mu_V^2 / \sigma_V^2$ and $\theta = \sigma_V^2 / \mu_V$, giving a Normal-Gamma distribution.
\end{itemize}

Note that when $\sigma_V^2 \rightarrow 0$ we have $\kappa_X \rightarrow 0$, which gives $g(v|\mu_V, \sigma_V) \rightarrow \delta (v - \mu_V)$, and reduces the symmetric NVM to a Gaussian distribution with variance $\mu_V$. Then, these functions can be used to probe deviations from Gaussianity in the Fourier coefficients of any noise process.

In this work, we want a distribution $g(v| \Lambda)$ that describes the statistics of the GWB spectrum, accounting for the Poisson shot noise and the properties of sources parametrized by astrophysical parameters $\Lambda$.

\section{Saddlepoint approximation for astrophysical population}

The average total power emitted by the population of GW sources can be written as the sum of all individual GW contributions \cite{Sesana_2008}. In characteristic strain units, it is given by

\begin{equation}
    h^2_c (f) = \int dz \int d\log \mathcal{M} \frac{d^3N}{dz d\log \mathcal{M} d\ln f_r} h^2 (f_r),
\end{equation}
given individual source contributions $h^2 (f)$ averaged over polarization and inclination

\begin{equation}
    h^2 (f) = \left( \frac{2}{5} \right) \left[ 4 \left ( \frac{G \mathcal{M}}{c^2}\right)^{5/3} \frac{(1 + z)^{2/3}}{d_M} \left( \frac{\pi f}{c} \right)^{2/3} \right]^2,
\label{eq:h2_polinc}
\end{equation}
with $\mathcal{M}$ the chirp mass, $z$ the redshift, $d_M$ the comoving distance to the binary and $f_r = (1 + z) f$ the rest-frame frequency of the binary.

The actual distribution of $h^2_c (f)$ is represented by a compound Poisson process where the number of sources contributing to each bin of the parameter space is an integer random variable following a Poisson distribution $\mathcal{P}(N)$ \cite{xue2025, holodeck}. This accounts for the shot noise due to the discreteness of the population. The discretized sum over $z_I$ and $ \log\mathcal{M}_J$ per log-frequency bin $\Delta \ln f = \Delta f / f$ gives

\begin{equation}
    h^2_c (f) = \sum_{IJ} \mathcal{P} \left \{ N (z_I, \log\mathcal{M}_J, f) \right \} \frac{h^2 (z_I, \log\mathcal{M}_J, f_r)}{\Delta f / f},
\label{eq:poisson_gwb}
\end{equation}
with the characteristic number of sources per bin $N (z_I, \log\mathcal{M}_J, f) = \Delta z \Delta \log\mathcal{M} \Delta \ln f \frac{d^3N}{dz d\log\mathcal{M} d \ln f_r} \propto f^{-11/3}$ for GW-driven circular binaries. In this work, the sum over $IJ$ is performed on a 50$\times$50 log-spaced grid of $\log_{10}z_I$ and $\log_{10}(\mathcal{M}_J / M_\odot)$ respectively between $[-2, 0.6]$ and $[7, 11]$.

Using the fact that the mean and variance of a Poisson distribution $\mathcal{P}(N)$ are respectively $N$ and $N$, we can easily find the mean $\mu_{h^2_c}$ and variance $\sigma^2_{h^2_c}$ of $h^2_c$ as

\begin{equation}
    \mu_{h^2_c} = \sum_{IJ} N (z_I, \log\mathcal{M}_J, f) h^2 (z_I, \log\mathcal{M}_J, f_r) \frac{f}{\Delta f},
\end{equation}

\begin{equation}
    \sigma^2_{h^2_c} = \sum_{IJ} N (z_I, \log\mathcal{M}_J, f) \left [ h^2 (z_I, \log\mathcal{M}_J, f_r) \frac{f}{\Delta f} \right ]^2.
\end{equation}

The latter can be used directly to perform moment matching with the distributions and parametrization given in \autoref{sec:NVM}.

In reference \cite{xue2025}, they use the definition of the cumulant generating function (CGF) for the compound Poisson process $h^2_c (f)$ that is given by

\begin{equation}
    K(is) = \sum_{IJ} N (z_I, \log\mathcal{M}_J, f)  \left [ e^{ish^2 (z_I, \log\mathcal{M}_J, f) \frac{f}{\Delta f}} - 1\right],
\end{equation}
and is related to the characteristic function $\varphi (s)$, i.e. the Fourier transform of the probability density $p(h^2_c)$ through $\varphi (s) = e^{K(is)}$.

We can find the distribution of $h^2_c$ by taking the inverse Fourier transform of the characteristic function \cite{satopolito_2024, xue2025}

\begin{equation}
    p(h_c^2) = \int ds e^{K(is) + ish_c^2} \approx \frac{e^{K(\hat s) - \hat s h_c^2}}{\sqrt{2 \pi K'' (\hat s)}},
\label{eq:saddlepoint}
\end{equation}
and this integral can be approximated using Laplace's method around the stationary point $\hat s$ defined by the equation $K'(\hat s) = h_c^2$. This is referred to as the saddlepoint approximation method \cite{saddlepoint}. We can solve numerically for any $h_c^2$ the equation $\log K'(\hat s) = \log h_c^2$ in log-scale using Newton's method since we know all analytical derivatives of the CGF.

\begin{figure}[h]
    \centering
    \includegraphics[width=\columnwidth]{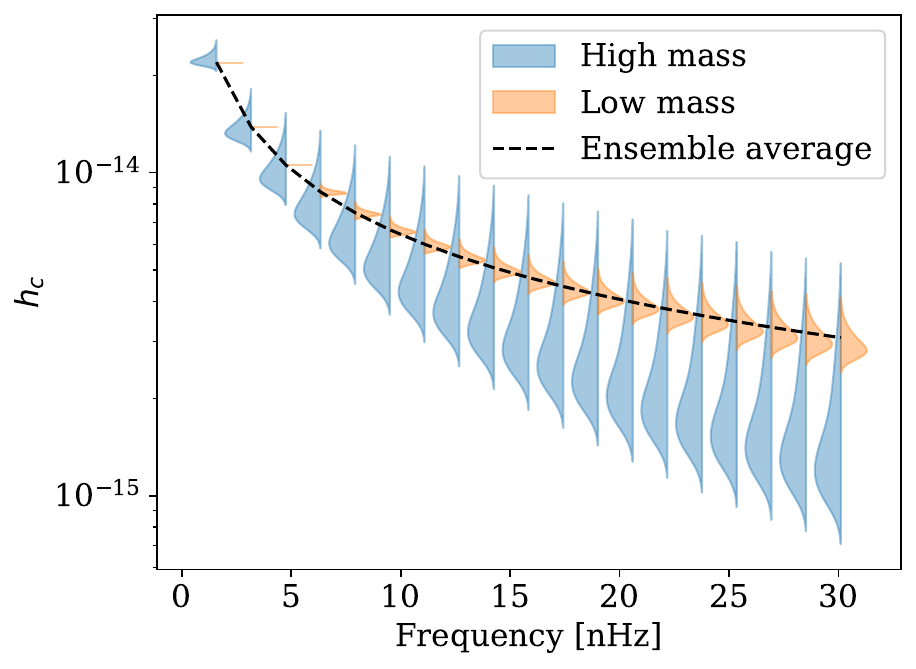}
    \caption{Saddlepoint approximation for the characteristic strain $h_c$ of the high mass and low mass population of \autoref{tab:models}. Both models produce the same average $h_c$. The high mass population $h_c$ fluctuates significantly more than the low mass.}
    \label{fig:spectrum}
\end{figure}

The advantage of this method is that it gives a relatively quick approximation of the real distribution from the CGF without requiring heavy training or precomputation. It only requires a model of the distribution of sources in chirp mass and redshift. The bottleneck is having to solve the equation $K'(\hat s) = h_c^2$ for each $h_c^2$.

\begin{figure}[h]
    \centering
    \includegraphics[width=\columnwidth]{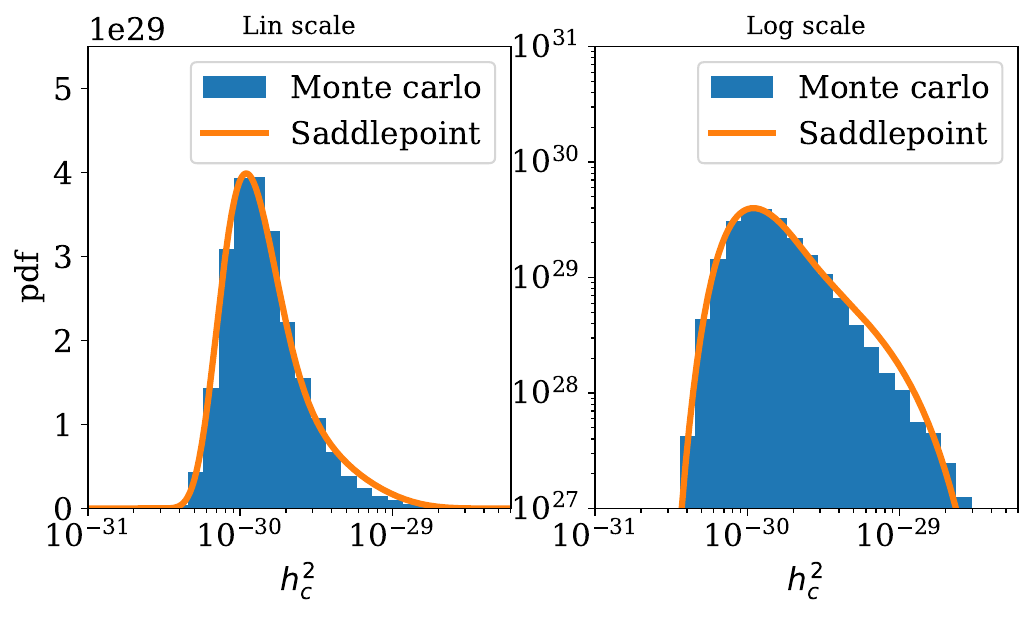}
    \caption{Compound poisson Monte Carlo versus saddlepoint approximation probability distribution. Most of the bulk distribution is correctly modeled, but we see in log scale that the approximation does not perfectly capture the tail behavior.}
    \label{fig:saddlepoint}
\end{figure}

In \autoref{fig:spectrum} we show the distribution of $h_c$ obtained for the models presented in \autoref{tab:models} and the next section. The violin plots show the cosmic variance given a population, and one realization of the GWB corresponds to a random draw from the violin plots at each frequency (equivalent to random draw from \autoref{eq:poisson_gwb}). In \autoref{fig:saddlepoint}, we overplot this distribution evaluated at $f=1/(yr)$ for the high mass model with a histogram generated from random draws of \autoref{eq:poisson_gwb}. We can see that the saddlepoint approximation catches most of the bulk distribution shape, but the log-scale plot highlights the slight difference at the high $h^2_c$ tail. We evaluate the goodness of the saddlepoint approximation with respect to the random draws from \autoref{eq:poisson_gwb} by computing the Hellinger distance between the two distributions in Appendix~\ref{app:hellinger}.

\section{Model and likelihood}

\subsection{Model}
\label{sec:astro_model}

To construct an astrophysical model we need to define $\frac{d^3N}{dz d\log_{10}\mathcal{M} d\ln f}$ that gives the distribution in redshift, chirp mass and frequency of the sources. More elaborate models might include mass ratio or eccentricity but we do not take into account these effects here \cite{Bonetti_2024, laal2024}. We define the number of sources per unit redshift $z$, chirp mass $\mathcal{M}$ and frequency $f$ as

\begin{equation}
    \frac{d^3N}{dz d\log_{10}\mathcal{M} d\ln f} = \frac{d^2n}{dz d\log_{10}\mathcal{M}} \frac{dV_c}{dz} \frac{dz}{dt_r} \frac{dt_r}{d \ln f_r},
\end{equation}
with
\begin{equation}
    \frac{dV_c}{dz} \frac{dz}{dt_r} = (1 + z) 4 \pi c d^2_M,
\end{equation}
that is given by cosmology and accounts for how the comoving volume $V_c$ changes with redshift. The frequency evolution term is given by the binary evolution due to GW emission. In this work, we only focus on circular GW driven binaries where

\begin{equation}
    \frac{dt_r}{d \ln f_r} = \frac{5}{96} \left(\frac{c^3}{G \mathcal{M}}\right)^{5/3} (\pi f_r)^{-8/3}.
\end{equation}

Still, accounting for more complex environmental effects is made possible in the code and the users are free to parametrize the frequency dependence as they wish. In essence, it would require a different frequency dependence than $f^{-8/3}$ \cite{satopolito2025anisotropy, raidal2026, Bonetti_2024}.

For the comoving number density of sources per unit volume, chirp mass and redshift, we use the phenomenological model used in \cite{Middleton_2015, satopolito2025anisotropy, raidal2026, Quelquejay_Leclere_Zombie}

\begin{equation}
    \frac{d^2n}{dz d\log_{10}\mathcal{M}} = \dot{n}_0 \frac{dt_r}{dz} \left ( \frac{\mathcal{M}}{10^7 M_\odot}\right )^{-\alpha} e^{-\mathcal{M} /\mathcal{M}_0} (1+z)^\beta e^{-z/z_0},
\end{equation}
where $\dot{n}_0$ is the comoving merger rate that is assumed to be constant.

This model is used as default for this work, but the code available at \url{https://github.com/mfalxa/zamari} can take any distribution and compute the associated statistics of the spectrum. It can be paired with outputs of semi-analytical models that provide more physically motivated population distributions \cite{bonoli2025}.

\subsection{Likelihood}

In PTA data analysis, we assume that the timing residuals are Gaussian and use a multivariate Gaussian likelihood that is constructed from the concatenated timing residuals of individual pulsars $\delta t = \cup_a \delta t_a$. In this work specifically, we consider a simplified model with only white noise and a GWB. For a Fourier decomposition of the GWB signal on a discrete Fourier basis $F$ with coefficients $\vec a$ (see \cite{van_Haasteren_2014} or the appendix of \cite{falxa_ng} for more mathematical details), the Gaussian likelihood is

\begin{equation}
\begin{aligned}
    \mathcal{L}(\delta t|\rho, \sigma) & = \int_{-\infty}^\infty d \vec a \mathcal{L}(\delta t - F \vec a|\sigma) \phi (\vec a | \rho)  \\
    & = \frac{\exp \left \{ -\frac{1}{2} \delta t^\top C^{-1} \delta t \right \}}{\sqrt{|2\pi C|}},
\end{aligned}
\label{eq:gaussian_likelihood}
\end{equation}
where $\phi (\vec a | \rho)$ is the Gaussian prior with variance $\rho$ enforcing that the Fourier coefficients are Gaussian distributed and the covariance matrix is given by

\begin{equation}
    C = \left[ (M \epsilon M^\top)_a + \sigma_{a, i}^2 \delta_{ij} \right] \delta_{ab} + \textrm{HD} (\zeta_{ab}) C_{\rm GWB},
\end{equation}
with $(M \epsilon M^\top)_a$ the timing model block obtained from the marginalization over first order errors of pulsar $a$ timing model \cite{van_Haasteren_2012, epta_timing} that removes power at low frequencies, $\sigma_{a, i}^2$ the white noise level (measurement errors) for the $i$th measurement of pulsar $a$ and $C_{GWB}$ the covariance matrix associated to the GWB signal. $C_{GWB}$ is described as a low frequency red noise with a covariance matrix that is the low rank approximation of the Wiener Khinchin integral at frequencies $f_k$ \cite{van_Haasteren_2014_cov}

\begin{equation}
    C_{\rm GWB} (t_i - t_j)= \sum_k \rho_k \cos \left( 2\pi f_k |t_i - t_j| \right),
\end{equation}
where $\rho_k$ is defined by the PSD and characteristic strain using $f S_h (f) = h^2_c (f)$ \cite{Romano_2017} which gives \footnote{In \autoref{eq:cross_correlated_residuals}, we have by definition of the expected value of the squared Fourier coefficients per unit frequency $\langle c_{00}^W \rangle/ (12 \pi^2 f^2) = S_{\delta t} (f)$ and $c_{00}^W = S_h(f)$.}

\begin{equation}
    \rho_k = S_{\delta t} (f_k) \Delta f = \frac{h^2_c (f_k)}{12 \pi^2 f^3 T},
\end{equation}
where $S_{\delta t} (f)$ is the one-sided PSD in timing residuals unit, $S_h (f)$ is the one-sided PSD in strain unit and we use $\Delta f = 1/T$ for equally spaced frequencies $f_k$ where $T$ is the total time of observation.

In the previous section, we presented the symmetric NVM distribution that allows to construct non-Gaussian priors by setting the correct hierarchical prior on $\rho$. Thus, for an NVM distribution parametrized by parameters $\Lambda$, using \autoref{eq:gaussian_likelihood} and \autoref{eq:nongaussian_prior}, we have

\begin{equation}
\begin{aligned}
    \mathcal{L}(\delta t|\Lambda, \rho, \sigma) & = \int_{-\infty}^\infty d \vec a \mathcal{L}(\delta t - F \vec a|\sigma) \phi (\vec a | \rho) g (\rho | \Lambda)\\
    & = \mathcal{L}(\delta t| \rho, \sigma) g (\rho | \Lambda),
\end{aligned}
\label{eq:ng_likelihood}
\end{equation}
which effectively corresponds to a non-Gaussian likelihood, targeting the free spectrum coefficients $\rho$ as the non-Gaussian components. When $g (\rho | \Lambda) \rightarrow \delta ( \rho - \rho_0)$, the likelihood reduces to the Gaussian case with free spectrum bins $\rho_0$. Sampling both $\rho$ and $\Lambda$ is equivalent to numerically performing the integral over $\rho$ as in \autoref{eq:nongaussian_prior}.

The posterior distribution is constructed from the product of the likelihood and prior probability distributions 

\begin{equation}
    p(\rho, \Lambda, \sigma | \delta t) \propto \mathcal{L}(\delta t | \rho, \sigma) g (\rho | \Lambda)  \pi (\Lambda) \pi (\sigma),
\label{eq:posterior}
\end{equation}
where $g (\rho | \Lambda)$ is given by the model and saddlepoint approximation defined in the previous section, $\pi (\Lambda)$ the prior probability on hyperparamters $\Lambda$ and $\pi (\sigma)$ the prior probability on all other noise parameters of the model $\sigma$.

Therefore, we can directly target the non-Gaussianity of any noise component from the Gaussian Likelihood by setting the appropriate hyperprior on the free spectrum parameters $\rho$. This is, in essence, equivalent to what \cite{xue2025} proposes as the Gaussian convolution, if we marginalize over the parameters $\rho$. In \cite{Quelquejay_Leclere_2023, Lamb_2023}, a procedure to fit $\Lambda$ from the marginalized distribution of $\rho$ obtained from the Gaussian likelihood is presented. The previous expression for the posterior probability distribution marginalized over $\rho$ can be rearranged as

\begin{equation}
    p(\vec \Lambda | \delta t) \propto \prod_{k=1}^{N_f} \int d \rho_k p(\rho_k | \delta t) g (\rho_k | \vec \Lambda) \pi (\vec \Lambda),
\end{equation}
assuming that frequencies $f_k$ are uncorrelated. This new expression of the likelihood is easily parallelizable, computationally less demanding and was already used in numerous publications. This is the one we use to produce results in \autoref{sec:astro_gwb}. The priors on parameters $\Lambda$ are given in \autoref{tab:priors}.

\section{Results}

We divide this section into two parts. First, we show how moment-matched distributions can be used to parameterize deviations from Gaussianity. We then compare the performance of the saddlepoint approximation and the moment-matched distribution in inferring the astrophysical parameters of the phenomenological model introduced \autoref{sec:astro_model}.

\subsection{Parametrized non-Gaussianity}
\label{sec:param_NG}

We simulate an ideal PTA with $N_{p} = 60$ pulsars uniformly distributed in the sky with a white noise level of $\sigma_a = 10^{-7}$s and a total time of observation $T = 20$ years. We inject a common HD correlated red noise to all pulsars on a grid of 30 equally spaced frequencies $f_k = k / T$ with $k$ an integer between 1 and 30. The vector of correlated Fourier coefficients for each pulsar $\vec{a}$ at $f_k$ are drawn from a multivariate Gaussian distribution $\mathcal{N}(\vec 0, \rho_k \chi_{ab})$ with $\chi_{ab}$ the $N_{p} \times N_{p}$ matrix containing HD correlation coefficients and $\rho_k$ is drawn from $g(\rho| \mu_\rho, \sigma_\rho)$ an inverse Gaussian distribution\footnote{The probability density $f(a) = \int_0 ^\infty d\rho \phi(a|\rho) g(\rho)$ is then a Normal-Inverse Gaussian distribution and has a well studied closed-form expression.}. Following the moment matching procedure in \autoref{eq:match_kurtosis} and \autoref{eq:ig}, the mean $\mu_\rho$ and standard deviation $\sigma_\rho$ of $g(\rho_k|\mu_\rho, \sigma_\rho)$ are chosen so that the auto-correlated PSD $S(f) = \mu_\rho$ and excess kurtosis $\kappa (f) = 3 \sigma^2_\rho$ of the injected signal in one pulsar are parametrized as

\begin{equation}
\begin{aligned}
    S (f) & = A f^{-\gamma},\\
    \kappa (f) & = 3 S^2 (f) \left ( \frac{f}{f_*}\right)^{11/3},
\end{aligned}
\end{equation}
with $A$ the amplitude, $\gamma$ the spectral index, and $f_*$ a characteristic frequency where non-Gaussianities become stronger, mimicking an astrophysical GWB with decreasing number of sources at higher frequencies.

The cross PSD and excess kurtosis can be found using Isserli's theorem

\begin{equation}
\begin{aligned}
    S_{ab} (f) & = A f^{-\gamma} \chi_{ab},\\
    \kappa_{abcd} (f) & = \left(\chi_{ab}\chi_{cd} + \chi_{ac}\chi_{bd} + \chi_{ad}\chi_{bc} \right) S_{aa}^2 (f) \left ( \frac{f}{f_*}\right)^{11/3},
\end{aligned}
\end{equation}
which reduces to the auto-correlated case for $a=b=c=d$.

We sample the posterior probability distribution of \autoref{eq:posterior} using the sampler \texttt{numpyro} \cite{phan2019} and the library \texttt{discovery} \cite{vallisneri_2025} to compute the Gaussian free spectrum likelihood $\mathcal{L}(\delta t | \rho, \sigma)$. We set an inverse Gaussian hyperprior on $\rho_k$ at each frequency $f_k$ as $g (\rho_k | A, \gamma, f_*)$ to control non-Gaussianties. The priors on hyperparameters are uniform with bounds $\log_{10} A = [-18, -11]$, $\gamma = [0, 7]$ and $\log_{10} f_* = [-9, -7]$.

\begin{figure}[h]
    \centering
    \includegraphics[width=\columnwidth]{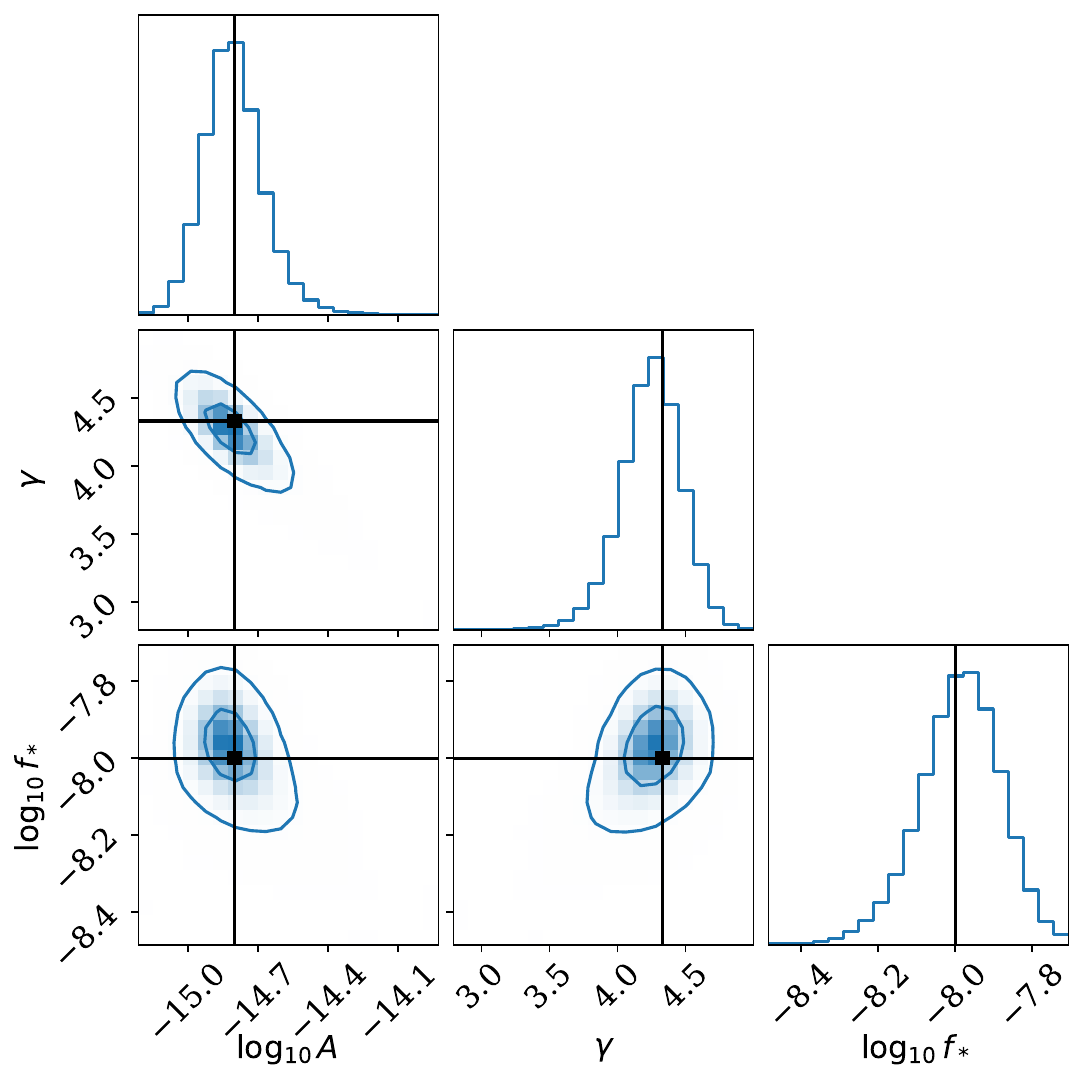}
    \caption{Posterior distribution for the simplistic GWB model that is used as an example. The injected signal had parameters $\log_{10} A = -14.8$, $\gamma=13/3$ and $\log_{10} (f_* / \rm Hz) = -8$.}
    \label{fig:corner_example}
\end{figure}

In \autoref{fig:corner_example}, we show the posterior distributions obtained for an injected signal with parameters $\log_{10} A = -14.8$, $\gamma=13/3$ and $\log_{10} (f_* / \rm Hz) = -8$, showing that the correct parameters are recovered. This is in essence similar to the powerlaw spectrum GWB searches that have been performed by PTA collaborations \cite{epta_wm3, ng15yr, ppta_dr3, cpta, mpta}, except here, deviations from Gaussianity are allowed and parameterized through the parameter $f_*$ that controls the excess kurtosis. Still, the spectrum of a realistic GWB does not follow an inverse Gaussian distribution and require more elaborate calculations \cite{satopolito_2024, xue2025}. In the following subsection, we compare the saddlepoint approximation versus a log-Normal distribution with mean and variance matched to the expected theoretical predictions.

\subsection{Astrophysical background}
\label{sec:astro_gwb}

We simulate an ideal PTA with 60 pulsars uniformly distributed in the sky with a white noise level of $\sigma_a = 10^{-7}$s with 20 years of observation. We inject an HD correlated GWB with a spectrum drawn for the Poisson process defined in \autoref{eq:poisson_gwb}. We study two kinds of population (i) a massive population with lower merger rate, inducing stronger spectral variance (ii) a lighter population with higher merger rate inducing weaker spectral variance. We compare each case with a moment matched log-normal distribution. The two populations are built to have the same average characteristic strain of $h_c (f) = 3 \times 10^{-15} (f / f_{yr})^{-2/3}$ (with $f_{yr} = 1 / (1yr)$) as in \autoref{fig:spectrum} and are generated using the parameters shown in \autoref{tab:models}.

\begin{table}[]
\centering
\begin{tabular}{ c | c | c }
  & High mass & Low mass \\  
 \hline
 $\log_{10} \dot{n}_0$ & -2.9 & -1.74 \\ 
 $\alpha$ & 0.5 & 0.5 \\  
 $\log_{10} \mathcal{M}_0$ & 9.3 & 8.3 \\
 $\beta$ & 0.5 & 0.5 \\
 $z_0$ & 1 & 1 \\
\end{tabular}
\caption{Parameters used for the high mass and low mass populations. They both give a GWB with an average characteristic strain $h_c = 3 \times 10^{-15} (f / f_{yr})^{-2/3}$ but different higher order moments.}
\label{tab:models}
\end{table}

\begin{table}[]
\centering
\begin{tabular}{ c | c }
 $\log_{10} \dot{n}_0$ & [-5, 0] \\ 
 $\alpha$ & [0, 3] \\  
 $\log_{10} \mathcal{M}_0$ & [7, 10] \\
 $\beta$ & [0, 3] \\
 $z_0$ & [0.1, 3] \\
\end{tabular}
\caption{Uniform prior ranges used for the population parameters.}
\label{tab:priors}
\end{table}

\subsubsection{High mass population}

The high mass population has a characteristic mass of $\log_{10} \mathcal{M}_0 = 9.3$. Then, some individual binaries are quite bright and few binaries are required to produce a signal with amplitude $h_c (f) = 3 \times 10^{-15} (f / f_{yr})^{-2/3}$. This tends to increase the spectral variance and boost the non-Gaussian nature of the signal.

\begin{figure}[h]
    \centering
    \includegraphics[width=\columnwidth]{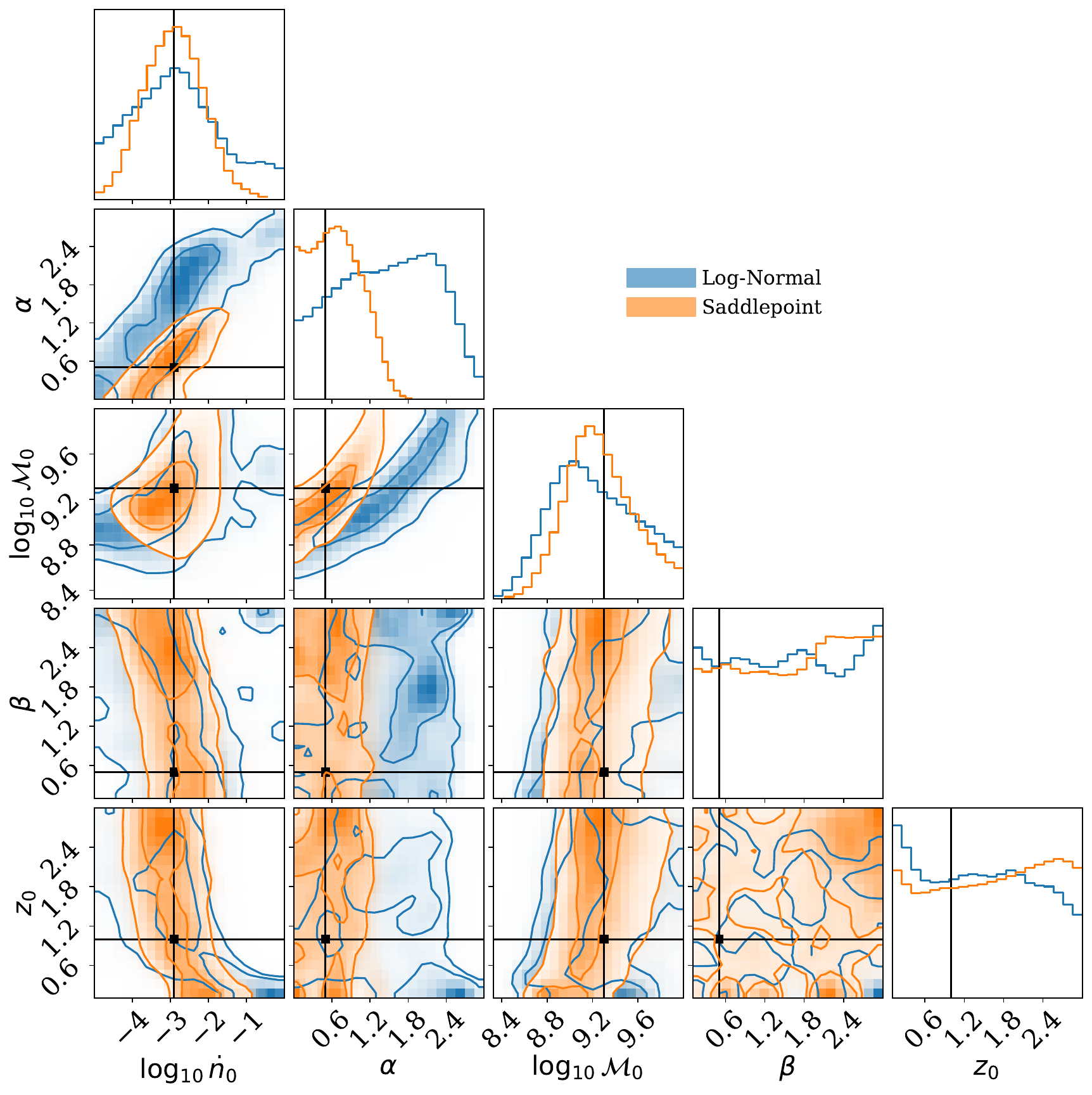}
    \caption{Posterior distributions of population hyperparameters for both log-Normal prior and saddlepoint approximation prior in the high mass population case.}
    \label{fig:corner_93}
\end{figure}

From the corner plot in \autoref{fig:corner_93}, we see that the saddlepoint distribution performs better than the moment matched log-normal distribution. Specifically for the parameters characterizing the mass function. This is because matching only the first two moments (mean and variance) does not ensure that higher order moments will accurately capture the higher $h^2_c$ tails of the distribution. Even though the saddlepoint method is an approximation, it still captures most of the bulk features and tail of the distribution that are crucial for precise inference. Still, even in a very ideal configuration of PTA, the posterior uncertainties on the parameters remain large. Specifically for $\beta$ and $z_0$ parameters that seem unconstrained.

\subsubsection{Low mass population}

The low mass population has a characteristic mass of $\log_{10} \mathcal{M}_0 = 8.3$. Then, the individual binaries are less bright and many binaries are required to produce a signal with amplitude $h_c (f) = 3 \times 10^{-15} (f / f_{yr})^{-2/3}$. This tends to Gaussianize the signal and reduce the spectral variance.

\begin{figure}[h]
    \centering
    \includegraphics[width=\columnwidth]{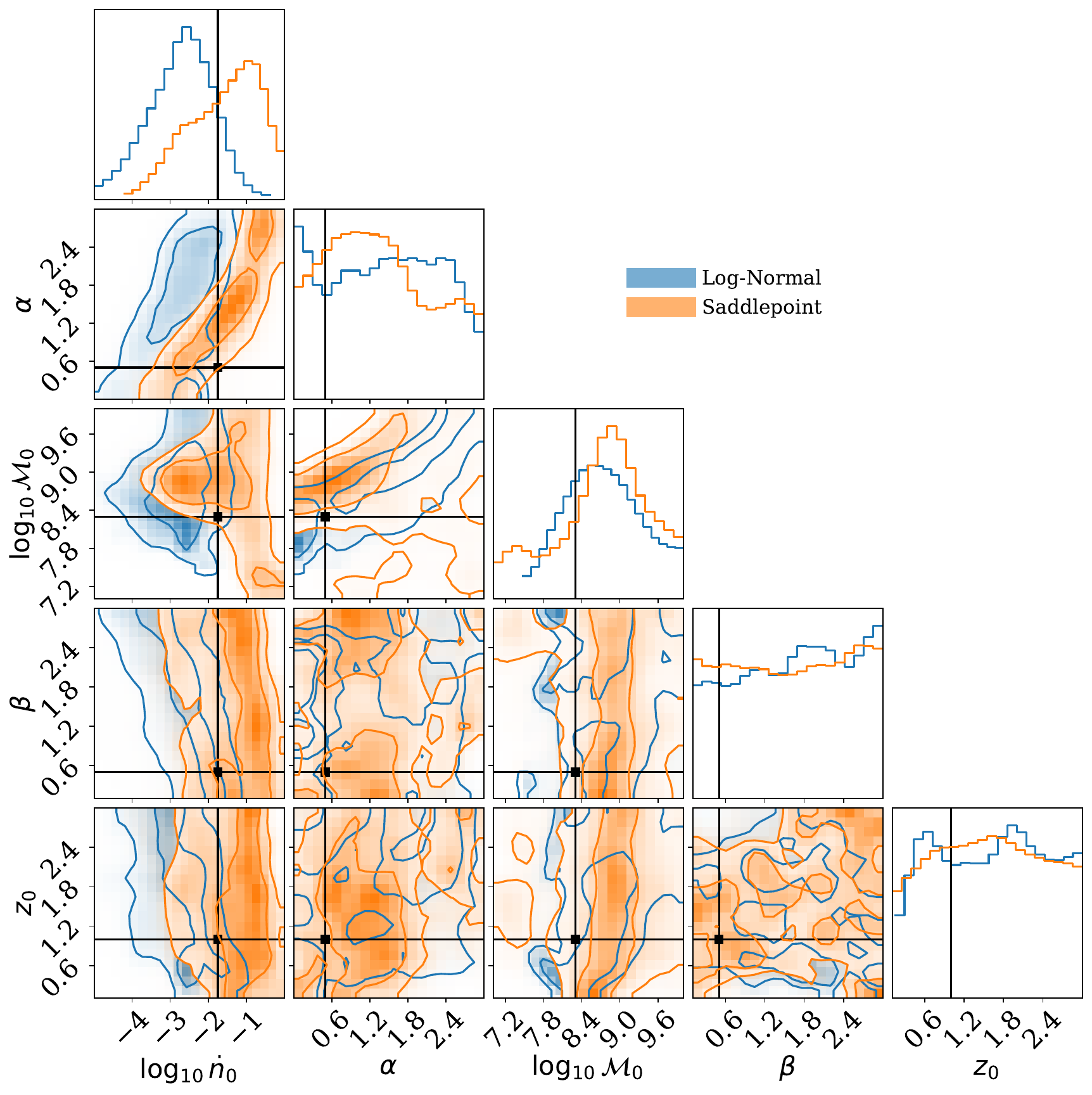}
    \caption{Posterior distributions of population hyperparameters for both log-Normal prior and saddlepoint approximation prior in the high mass population case.}
    \label{fig:corner_83}
\end{figure}

In \autoref{fig:corner_83}, the moment matched log-normal and saddlepoint approximation perform equally within posterior uncertainties, because in this case, the spectral variance is small, hence the log-normal and saddlepoint distributions are more similar in shape. Not modeling the higher order moments and high $h^2_c$ tails has a lesser impact since the GWB is nearly Gaussian.

\section{Conclusion}

We have presented a flexible method to include non-Gaussian behaviors in the PTA likelihood and approximate the statistics of a GWB for any given population model. The method is based on the saddlepoint approximation that was first presented in \cite{saddlepoint} and also mentioned recently in the context of PTA in \cite{goncharov2026}. The main goal of this work is to provide a ready-to-use tool that gives quick estimates of the distribution of characteristic strain $h_c$ for any given model. This enables Bayesian parameter inference or simply more realistic representations of the expected signal for other scientific publications that usually only plot the expected mean characteristic strain $h_c$. Knowing the distribution of $h_c$ allows to model the non-Gaussian statistics of the GWB through Gaussian scale mixture distributions.

We have shown that the approximation closely follows the true expected distribution that is obtained by sampling a compound Poisson process and computing the Hellinger distance between the two distributions. It seems that the saddlepoint approximation method performs equally as previously introduced machine learning based density estimators using normalizing flows \cite{laal2024}. That said, machine learning techniques could still allow more complex modeling (including eccentricity for example). Their real bottleneck lies in the number of simulations required for an accurate prediction of the distribution.

We test the parameter inference of a specific phenomenological model by performing simulations of ideal PTA datasets and Bayesian recovery of the posterior distributions. The saddlepoint method is compared to a moment-matched log-normal distribution for which the mean and variance are exactly those expected analytically from the population. However, an incorrect modeling of higher order moments leads to errors in the inference that can bias the recovery. We compare the results between a low mass and high mass population, that are, respectively, strongly non-Gaussian and more Gaussian (i.e. less spectral fluctuations), and show that high mass populations give more information about the SMBHB mass function through stronger spectral fluctuations.

Including the real statistics of the GWB in PTA analyses is essential to capture the astrophysical nature of the signal. Accessing deeper information about the mass function or number of SMBHB is what can enable proper astrophysical population studies with GWs and enable multi-messenger investigations. In future works, the inclusion of eccentricity will be become crucial. These features are usually not accounted for in standard PTA inference, but they might carry very interesting information about the demographics of SMBHBs.

\section*{Acknowledgements}

MF thanks Silvia Bonoli, Chiara Cecchini, Gabriela Sato-Polito, Xiao Xue, David Izquierdo-Villalba, Mauro Pieroni, Hippolyte Quelquejay Leclere, Alberto Sesana and everyone at the GGI Listening to the Cosmos: New Frontiers in Gravitational Wave Physics 2026 for sharing their code, knowledge and interesting discussions. The jaxification of the code was performed with the help of Claude code. MF acknowledges support from the Spanish Ministerio de Ciencia e Innovaci\'on through project PID2024-159201NB-C21.

\appendix

\section{Overlap reduction function}
\label{app:orf}

The fully developed expression for the cross response between pulsars $a$ and $b$ is

\begin{equation}
\begin{aligned}
    & R_a^* (\hat \Omega, f_s) R_b (\hat \Omega, f_s) \\
    &= \left[1 - e^{2i\pi f_s L_a (1 + \hat \Omega \cdot \hat p_a )} - e^{-2i\pi f_s L_b (1 + \hat \Omega \cdot \hat p_b )} \right.\\
    & \left.+ e^{2 i \pi f_s (L_a (1 + \hat \Omega \cdot \hat p_a ) -  L_b (1 + \hat \Omega \cdot \hat p_b))} \right] \\
    & \times \left [ \frac{1 + \cos ^2 \iota}{2} F_a^+ - i \cos \iota F_a ^\times \right ]\\
    & \times\left[ \frac{1 + \cos ^2 \iota}{2} F_b^+ + i \cos \iota F_b ^\times \right],
\end{aligned}
\end{equation}

In \cite{Anholm_2009, Romano_2017} it is shown that in the long arm limit $f_s L >> 1$, the first rapidly oscillating frequency dependent term gets damped when integrated over the sky, because integrating the exponential yields a term $\sim fL$ in the denominator. Therefore, it is a reasonable approximation to consider it to be $\approx 1$ and the response to be frequency independent, except when $a=b$, because the last exponential term is also 1, which accounts for the pulsar term present in the timing residuals. Therefore we have

\begin{equation}
    \int \frac{d \hat \Omega}{4\pi} R_a^* (\hat \Omega, f_s) R_b (\hat \Omega, f_s) \approx (1 + \delta_{ab}) \int \frac{d \hat \Omega}{4\pi} R_a^* (\hat \Omega) R_b (\hat \Omega),
\end{equation}
where $\delta_{ab}$ is accounting for the autocorrelated pulsar term.

We can calculate the polarization and inclination averaged response as

\begin{equation}
\begin{aligned}
    & \langle R^*_a(\hat \Omega) R_b(\hat \Omega) \rangle_{\psi, \iota} \\
    & = \int \frac{d \cos \iota}{2} \frac{d \psi}{\pi} \left [\left(\frac{1 + \cos^2 \iota}{2} \right)^2 [ (F^+_a F^+_b \cos^2 2\psi \right.\\
    & + F^\times_a F^\times_b \sin^2 2\psi - \cos 2\psi \sin 2\psi (F^+_a F^\times_b + F^\times_a F^+_b)]\\
    & + \left( \cos \iota \right)^2 [ F^+_a F^+_b \sin^2 2\psi +  F^\times_a F^\times_b \cos^2 2\psi \\
    &  + \cos 2\psi \sin 2\psi (F^+_a F^\times_b + F^\times_a F^+_b)\\
    & \left. + i\left(\frac{1 + \cos^2 \iota}{2} \right)\left( \cos \iota \right) (-F^+_a F^\times_b + F^\times_a F^+_b)] \right]\\
    & = \frac{2}{5} \left [ F^+_a F^+_b + F^\times_a F^\times_b \right].
\end{aligned}
\end{equation}

The imaginary part is an odd function of inclination that vanishes with the integration, as a consequence of assuming a smooth and uniform distribution of inclination and polarization. The prefactor $2/5$ goes into the definition of the polarization and inclination averaged strain amplitude in \autoref{eq:h2_polinc}

Then, combining the previous and omitting the 2/5 factor, we find

\begin{equation}
\begin{aligned}
    \Gamma_{00,ab} & \equiv (1 + \delta_{ab}) \int \frac{d \hat \Omega}{4\pi}\left [ F^+_a(\hat \Omega) F^+_b(\hat \Omega) + F^\times_a(\hat \Omega) F^\times_b(\hat \Omega) \right ]\\
    & = \frac{2}{3} \textrm{HD} (\zeta_{ab}).
\end{aligned}
\end{equation}
which by definition is the Hellings-Down correlation pattern $\textrm{HD} (\zeta_{ab})$ for an unpolarized and isotropic GWB, with $\zeta_{ab}$ the angle between pulsar $a$ and $b$ in the sky. The factor $2/3$ comes from the overlap integral $\int (d \hat{\Omega} / 4\pi) [(F_a^+)^2 + (F_a^\times)^2] = 1/3$ when $a=b$ (as calculated in cite refs) giving $\Gamma_{00,aa} = 2/3$ and defining $\textrm{HD}(0) = 1$.

\section{Gaussian approximation}
\label{app:gaussian_approx}

We build a toy model of a compound Poisson process $S$ that is made of $N$ single jumps $h_i$. Each jump corresponds to a sine wave with fixed frequency $f$, a random phase $\Phi_i$ uniform between $0$ and $2\pi$ and random amplitude $\sqrt{h^2}$ that is given by the astrophysical model of sources. We have

\begin{equation}
    S = \sum_{i=1}^{N} h_i,
\label{eq:toy_compound}
\end{equation}
with $N \sim \mathcal{P}(\bar{N})$ a Poisson distributed variable, $\bar{N}$ being the average number of sources contributing at frequencies $f$, obtained by integrating $d^3 N / dz d\log_{10} \mathcal{M} d\ln f$ with respect to $z$ and $d\log_{10} \mathcal{M}$, and

\begin{equation}
    h_i = X_i \sqrt{h^2}.
\end{equation}

The important difference here is that the sum is performed in $h_i$, thus accounting for the phase of each jump and considering interference between sources \cite{xue2025}, contrary to the sum performed in $h_i^2$ that omits this effect. We want to compare two cases where $X_i$ is a single random phasor (real case) and $X_i$ is a single Gaussian jump, representing the Gaussian approximation. In essence, this corresponds to a Brownian motion that is the sum of individual Gaussian jumps for which the total sum $S$ is also Gaussian \cite{Quelquejay_Leclere_2026}. We focus on the real part of the signal and define

\begin{itemize}
    \item $X_{i,A} = \textrm{Re} \{e^{i \Phi_i}\} \sim \textrm{Arcsine}(-1, 1),$
    \item $X_{i,G} \sim \mathcal{N}(0, 1 / \sqrt{2}),$
\end{itemize}
ensuring that both individual jumps have a variance of $1/2$.

We define the amplitude distribution $h^2$ by using the fact that $p(h^2) \propto dN / dh^2$ which itself can be obtained through a change of variable between $\mathcal{M}$ and $h^2$ \cite{satopolito_2024} using \autoref{eq:h2_polinc}

\begin{equation}
    \frac{dN}{dh^2 d \ln f} = \int dz \frac{d^3N}{dz d \mathcal{M} d\ln f}\mathcal{M} \left | \frac{d \mathcal{M}}{d h^2} \right |,
\end{equation}

and integrating numerically over $z$ and numerically normalized. This quantity is proportional to the probability distribution of the amplitude of a single source randomly drawn from the population.

For a compound Poisson process as defined in \autoref{eq:toy_compound}, the characteristic function $\Phi_S(s)$, i.e. the Fourier transform of the probability density function of the sum $S$, is given by

\begin{equation}
    \Phi_S (s) = \exp \left \{ \bar{N} \left( \int dh^2 p(h^2) \Phi_X \left(s \sqrt{h^2} \right) - 1 \right )\right \},
\end{equation}
where $\Phi_X(s)$ is the characteristic function of the individual jump $X_i$ for each case

\begin{itemize}
    \item Arcsine $\Phi_{X,A} (s) = J_0(s)$, the Bessel function of the first kind
    \item Gaussian $\Phi_{X,G} (s) = \exp \{-s^2 / 4 \},$
\end{itemize}
yielding the associated compound Poisson characteristic functions $\Phi_{S,A}$ and $\Phi_{S,G}$.

The two individual jump characteristic functions $\Phi_{X,A} (s)$ and $\Phi_{X,G} (s)$ obviously have very different behaviors. They are quite similar around the bulk $s=0$ but the Bessel function strongly oscillates and slowly decays at higher $s$ while the Gaussian quickly decays without oscillating. This difference in asymptotic behavior can induce significant differences at the tails of the distributions.

We identify two sources of randomness (i) the Poisson fluctuation of the number of sources (ii) the astrophysical source amplitude variability. We want to quantify how both random contributions actually wash out most of the differences between the Gaussian and the Arcsine jump cases in $\Phi_S (s)$, hence validating the Gaussian ensemble approximation.

We compute the Euclidean distance between $p_A(S)$ and $p_G(S)$, respectively, the probability distribution of $S$ for the phasor jump and the Gaussian jump. Noting that $\Phi_{S,A} - \Phi_{S,G}$ is the Fourier transform of $p_A(S) - p_G(S)$, Parseval's identity gives

\begin{equation}
    \int dS |p_A(S) - p_G(S)|^2 = \frac{1}{2\pi} \int ds |\Phi_{h,A} - \Phi_{h,G}|^2.
\end{equation}

Finally, we define the relative difference with respect to $p_G(S)$ as

\begin{equation}
    (\Delta p)^2 = \frac{\int ds |\Phi_{S,A} - \Phi_{S,G}|^2}{\int ds |\Phi_{S,G}|^2}.
\end{equation}

\begin{figure}[h]
    \centering
    \includegraphics[width=\columnwidth]{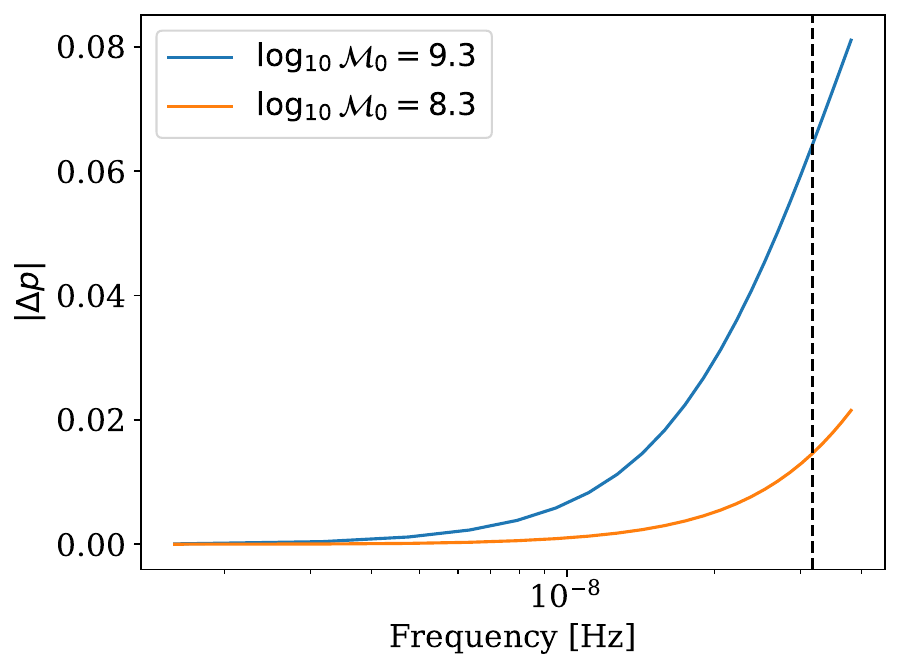}
    \caption{Relative difference between the arcsine and Gaussian jump models for two astrophysical populations (massive and light). The black dashed line shows the frequency of 1/(1yr).}
    \label{fig:delta_p}
\end{figure}

In \autoref{fig:delta_p}, we show values of $\Delta p$ evaluated at different frequencies $f$ for two models, one with high characteristic mass $\log_{10} \mathcal{M}_0 = 9.3$ and low characteristic mass $\log_{10} \mathcal{M}_0 = 8.3$. For the high mass case, the difference between the Gaussian and arcsine models can go up to $8 \%$ after $f=1/(yr)$ whereas for low masses only around $2 \%$. The high mass population will have a higher chance to produce resolved individual binaries for which the Gaussian approximation is no longer valid. This is what produces high amplitude tails in $p(S)$ that cannot be fully accounted for by the Gaussian approximation. Still, the difference is significantly suppressed by the introduction of Poisson fluctuations and amplitude variations. At low $\bar{N}$, the $p(h^2)$ weighted integral of individual jumps damps the oscillations of the Bessel function, while at high $\bar{N}$, the central limit theorem becomes valid. The combination of these two effects tend to reduce the difference between the models.

Nevertheless, this toy model assumes an unpolarized GWB and does not account for the GW response accross pulsars. Morevoer, the pulsar term is omitted, which might introduce additional interference between sources \cite{xue2025}. That said, it still provides a simple picture to understand why the Gaussian ensemble approximation can still be valid, as long as the Poisson and amplitude fluctuations are accounted for.

\section{Hellinger distance}
\label{app:hellinger}

We compute the Hellinger distance between a kernel density estimate (KDE) obtained with 5000 draws from the compound Poisson process of \autoref{eq:poisson_gwb} and the saddlepoint approximation obtained from \autoref{eq:saddlepoint}. The squared Hellinger distance defined as

\begin{equation}
    H^2(f, g) = \frac{1}{2}\int dx \left ( \sqrt{f(x)} - \sqrt{g(x)}\right )^2,
\end{equation}
gives a measure of the difference between two probability distributions.

We compute $H(f, g)$ for 1000 different parameter values from the priors in \autoref{tab:priors} and focus on the distribution of $h^2_c$ at $f=1 yr^{-1}$ where we expected a significant high $h^2_c$ tail in the distribution.

\begin{figure}[h]
    \centering
    \includegraphics[width=\columnwidth]{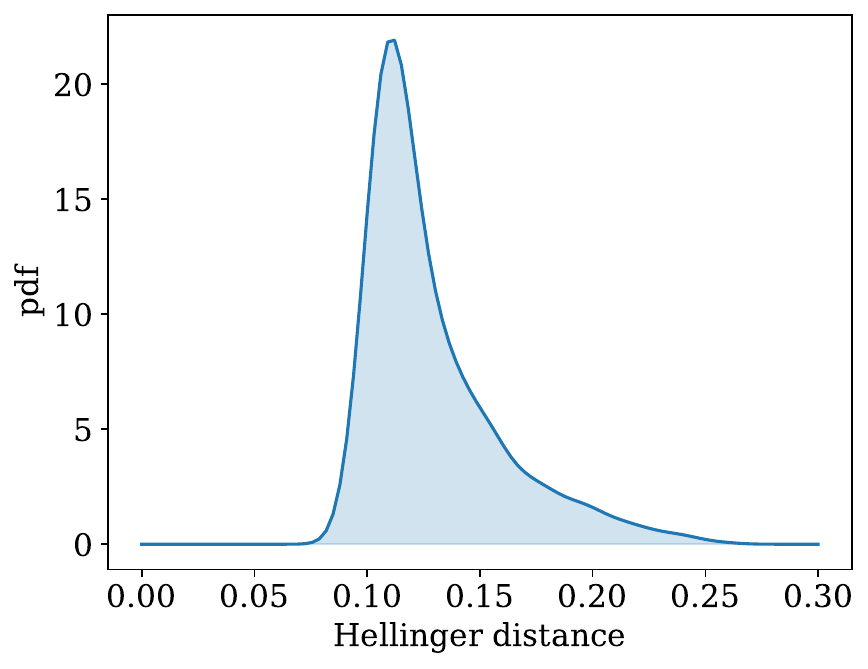}
    \caption{Hellinger distance between a KDE of compound poisson Monte Carlo draws against the saddlepoint approximation of the $p(h^2)$.}
    \label{fig:hellinger}
\end{figure}

In \autoref{fig:hellinger} we show the difference between the true distribution and the approximation. The saddlepoint approximation seems to perform as well as normalizing flow techniques \cite{laal2024} without requiring heavy training, which gives it a considerable advantage. However, the implementation of effects like eccentricity will require further development, as it is for now only accounted for thanks to machine learning techniques \cite{Bonetti_2024}.

\bibliography{saddlepoint.bib}
% Produces the bibliography via BibTeX.

\end{document}